\documentclass[12pt]{iopart}

\usepackage{graphicx}
\usepackage{subfigure}  
\begin{document}

\title[]{Study of Dynamical $K/\pi$ and $p/\pi$ Fluctuations in $\sqrt{s_{NN}}$ = 22.4 GeV Cu+Cu Collisions from the STAR Experiment}

\author{Terence J Tarnowsky (for the STAR Collaboration)}

\address{National Superconducting Cyclotron Laboratory, Michigan State University, East Lansing, MI 48824, USA}
\ead{tarnowsk@nscl.msu.edu}
\begin{abstract}

Results from the measurement of dynamical $K/\pi$ and $p/\pi$ ratio fluctuations are presented. Dynamical fluctuations in global conserved quantities such as baryon number, strangeness, or charge may be observed near a QCD critical point. The first measurement of results of $K/\pi$  and $p/\pi$ fluctuations from Cu+Cu collisions at an energy $\sqrt{s_{NN}}$ = 22.4 GeV are shown. This energy range is close to the top SPS energy and has garnered additional interest due to the proposed QCD critical point search at RHIC. The new Cu+Cu data is compared to previous results from Au+Au collisions at the energies $\sqrt{s_{NN}}$ = 19.6, 62.4, 130, and 200 GeV, and to measurements from the NA49 experiment. The dynamical $K/\pi$ fluctuation results from Cu+Cu at $\sqrt{s_{NN}}$ = 22.4 GeV are in good agreement with previous measurements from the STAR and NA49 experiments near that energy. Results are compared to theoretical predictions from several models including the HIJING and UrQMD models. 
\end{abstract}

\maketitle

\section{Introduction}

Fluctuations and correlations are well known signatures of phase transitions. In particular, the quark/gluon to hadronic phase transition may lead to significant fluctuations. As part of a proposed beam energy scan at the Relativistic Heavy Ion Collider (RHIC), the search for the QCD critical point will make use of the study of correlations and fluctuations, particularly those that could be enhanced during a phase transition that passes close to the critical point. Particle ratio fluctuations are an observable that has been studied as a function of energy and system-size. It will also provide additional information to assist in locating the QCD critical point when measured as part of the proposed beam energy scan.

Dynamic particle ratio fluctuations, specifically fluctuations in the $K/\pi$ and $p/\pi$ ratio, can provide information on the quark-gluon to hadron phase transition. These can be measured using the variable $\nu_{dyn}$, originally introduced to study net charge fluctuations \cite{nudyn1, nudyn2}. $\nu_{dyn}$ quantifies deviations in the particle ratios from those expected for an ideal statistical Poissonian distribution. The definition of $\nu_{dyn,K/\pi}$ (describing fluctuations in the $K/\pi$ ratio) is,

\begin{eqnarray}
\nu_{dyn,K/\pi} = \frac{<N_{K}(N_{K}-1)>}{<N_{K}>^{2}}
+ \frac{<N_{\pi}(N_{\pi}-1)>}{<N_{\pi}>^{2}}
- 2\frac{<N_{K}N_{\pi}>}{<N_{K}><N_{\pi}>}
\label{nudyn}
\end{eqnarray}

where $N_{K}$ and $N_{\pi}$ are the number of kaons and pions in a particular event, respectively. In this proceeding, $N_{K}$ and $N_{\pi}$ are the total charged multiplicity for each particle species. A formula similar to (\ref{nudyn}) can be constructed for $p/\pi$ fluctuations. $\nu_{dyn}$ = 0 for the case of a Poisson distribution of kaons and pions and is largely independent of detector acceptance and efficiency in the region of phase space being considered \cite{nudyn2}. An in-depth study of $K/\pi$ fluctuations in Au+Au collisions at $\sqrt{s_{NN}}$ = 200 and 62.4 GeV was carried out by the STAR experiment \cite{starkpiprl}.

Previous measurements of particle ratio fluctuations utilized the variable $\sigma_{dyn}$ \cite{NA49}, 

\begin{equation}
\sigma_{dyn} = sgn(\sigma_{data}^{2}-\sigma_{mixed}^{2})\sqrt{|\sigma_{data}^{2}-\sigma_{mixed}^{2}|}
\label{signudyn}
\end{equation}

where $\sigma$ is the relative width of the $K/\pi$ or $p/\pi$ distribution in either real data or mixed events. It has been shown that $\nu_{dyn}$ and $\sigma_{dyn}$ are mathematically related as $\sigma_{dyn}^{2} \approx \nu_{dyn}$.

The 2005 run at RHIC included Cu+Cu collisions at an energy of $\sqrt{s_{NN}}$ = 22.4 GeV. This was the second lowest energy run at RHIC with enough statistics to perform particle ratio fluctuation measurements. A previous run at $\sqrt{s_{NN}}$ = 19.6 GeV with Au+Au was also utilized to measure particle ratio fluctuations. However, because of the smaller system size of Cu+Cu, the collisions at $\sqrt{s_{NN}}$ = 22.4 GeV provide the lowest energy density presently available at RHIC to study particle ratio fluctuations. Collisions at $\sqrt{s_{NN}}$ = 22.4 GeV correspond to a baryon chemical potential ($\mu_{B}$) of 184 MeV, calculated from the parameterization \cite{Cleymans}, $\mu_{B}(\sqrt{s}) = \frac{1.308\: {\rm GeV}}{1 + 0.273\: {\rm GeV}^{-1}\sqrt{s}}$.
This is a range in $\mu_{B}$ space that will be covered by the forthcoming energy scan at RHIC. These measurements will provide a baseline for future studies of particle ratio fluctuations at all energies.

\section{Experimental Analysis}

The data presented here for $K/\pi$ and $p/\pi$ fluctuations was acquired by the STAR experiment at RHIC from minimum bias (MB) Cu+Cu collisions at a center-of-mass collision energy ($\sqrt{s_{NN}}$) of 22.4 GeV \cite{STAR}. The main particle tracking detector at STAR is the Time Projection Chamber (TPC) \cite{STARTPC}. A cut on the longitudinal position of the collision vertex ($v_{z}$) restricted events to within $\pm 30$ cm of the STAR TPC. All detected charged particles in the pseudorapidity interval $|\eta| < 1.0$ were measured. The transverse momentum ($p_{T}$) range for pions and kaons was $0.2 < p_{T} < 0.6$ GeV/c, and for protons was $0.4 < p_{T} < 1.0$ GeV/c. Charged particle identification involved measured ionization energy loss ($dE/dx$) in the TPC gas and total momentum ($p$) of the track. The energy loss of the identified particle was required to be less than two standard deviations (2$\sigma$) from the predicted energy loss of that particle at a particular momentum. A second requirement was that the measured energy loss of a pion/kaon was more than 2$\sigma$ from the energy loss prediction of a kaon/pion, and similarly for $p/\pi$ measurements. A third requirement was used to veto electron candidates. Identified pions, kaons, or protons with measured energy losses within 1$\sigma$ of predicted energy loss of an electron were rejected. The centralities used in this analysis account for 0-10, 10-20, 20-30, 30-40, 40-50, and 50-60\% of the total hadronic cross section.

\section{Results and Discussion}

\begin{figure}
\centering
\subfigure[]{
\includegraphics[scale=0.37]{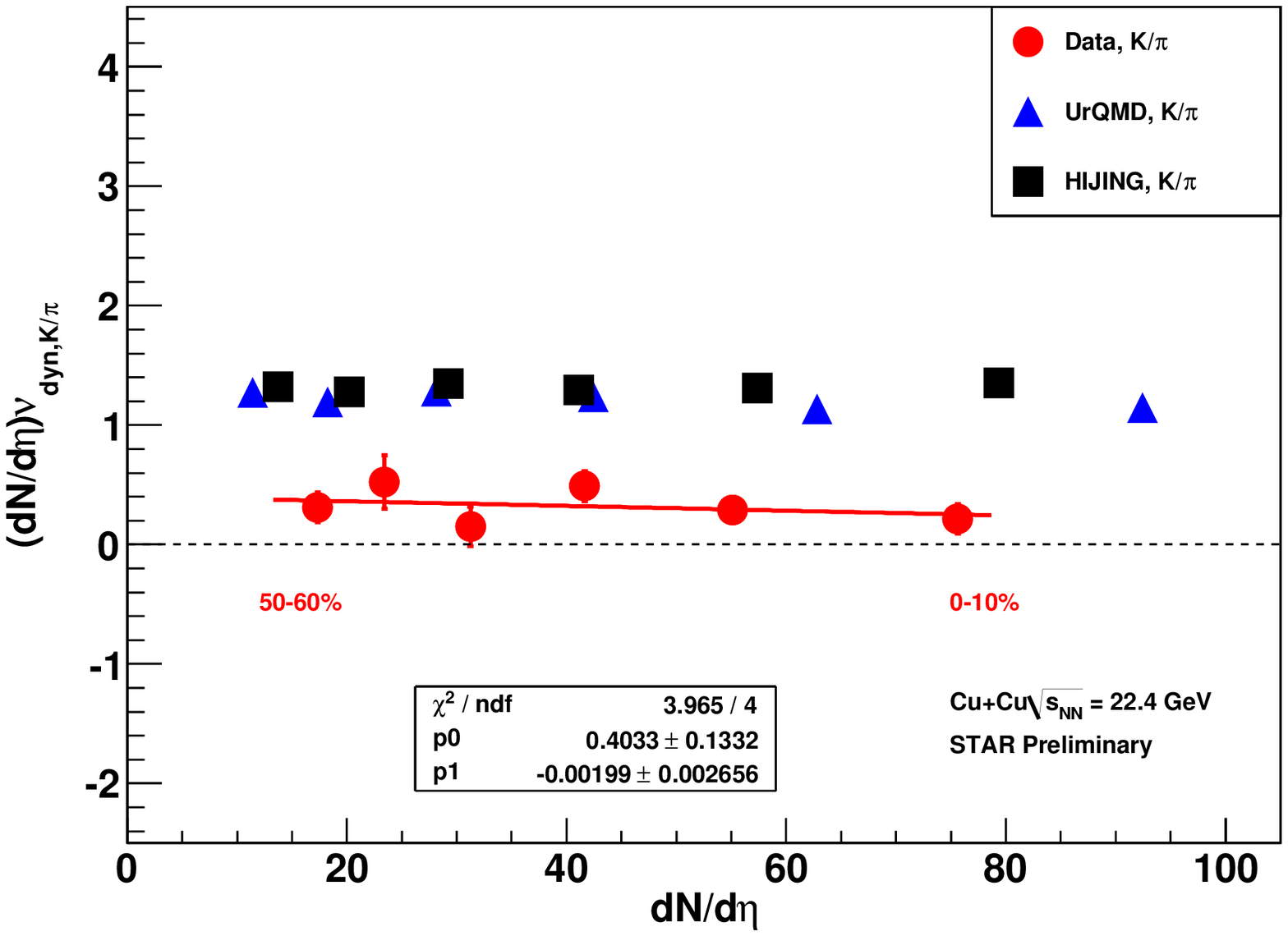} 
\label{nudynkpi}
}
\subfigure[]{
\includegraphics[scale=0.37]{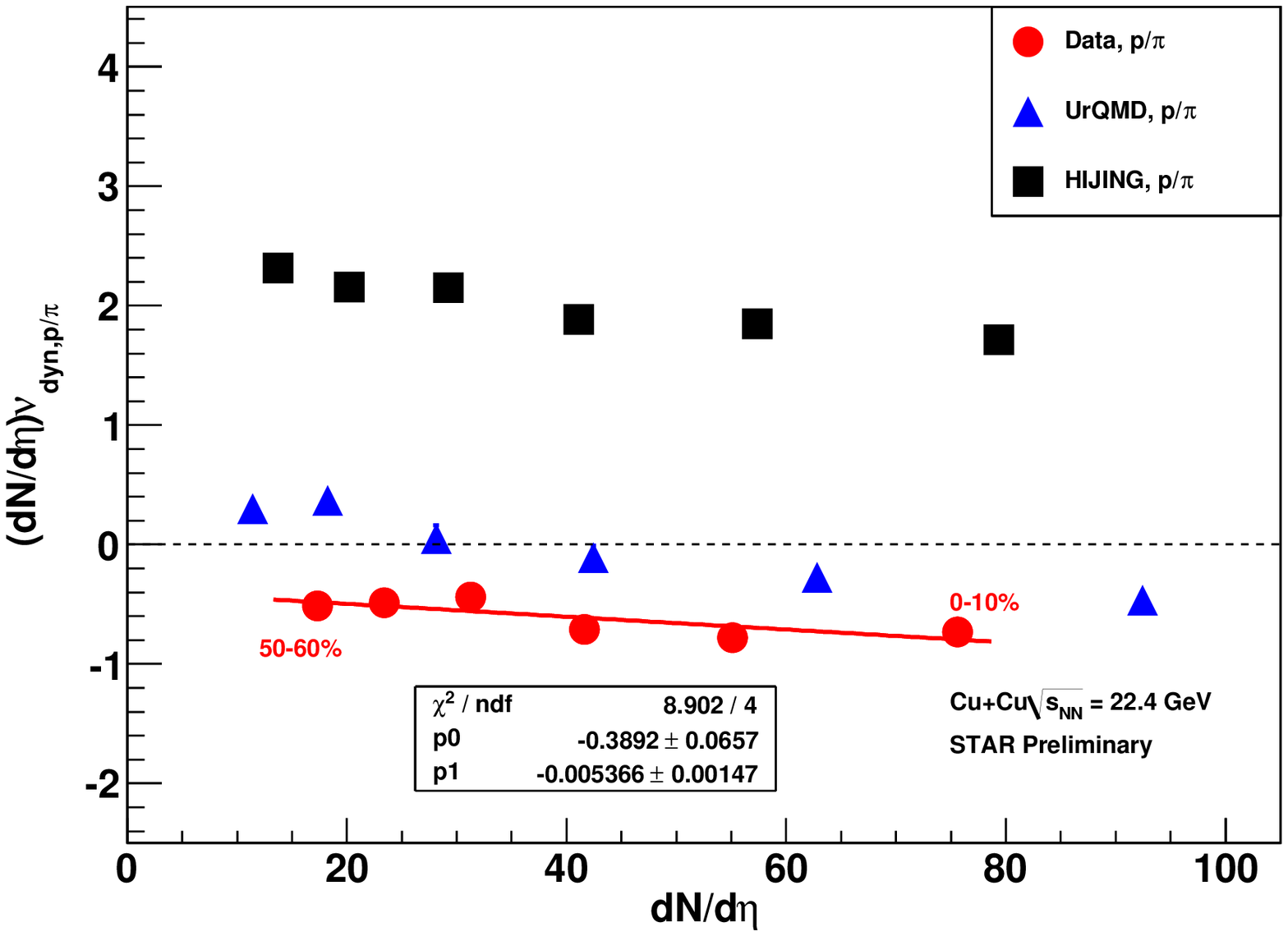} 
\label{nudynppi}
}
\caption{Results for $K/\pi$ (left) and $p/\pi$ (right) fluctuations as measured by $\nu_{dyn}$ from Cu+Cu collisions at $\sqrt{s_{NN}}$ = 22.4 GeV scaled by uncorrected charged particle multiplicity, $dN/d\eta$. The data (red circles) is compared to two models, UrQMD (blue triangles) and HIJING (black squares). A linear fit to the data and its parameters are also shown. All errors are statistical.}
\label{Fig1}
\end{figure}

The results from measurements of $K/\pi$ and $p/\pi$ fluctuations from Cu+Cu collisions at $\sqrt{s_{NN}}$ = 22.4 GeV are presented. Figure \ref{nudynkpi} shows the measured values of $\nu_{dyn}$ as a function of centrality for $K/\pi$ fluctuations scaled by the charged particle multiplicity, $dN/d\eta$ for the data (red circles) and two models, UrQMD \cite{UrQMD} (blue triangles) and HIJING \cite{HIJING} (black squares). The value of $dN/d\eta$ from the data is the uncorrected charged particle multiplicity. All displayed errors are statistical. A linear fit to the data is also shown. It was found in Au+Au collisions at both $\sqrt{s_{NN}}$ = 62.4  and 200 GeV that $\nu_{dyn,K/\pi}$ scales linearly with $dN/d\eta$ at low multiplicities \cite{starkpiprl}. Figure \ref{nudynkpi} demonstrates that $\nu_{dyn,K/\pi}$ also scales linearly in Cu+Cu collisions at $\sqrt{s_{NN}}$ = 22.4 GeV. The use of $dN/d\eta$ removes the 1/$N_{ch}$ dependence of $\nu_{dyn}$ and acts as an approximation for the size of the system. UrQMD and HIJING also reproduce this behavior and are in good agreement with each other, however they over predict the magnitude of the fluctuations compared to the data. The fluctuations are always positive, indicating the self-correlation terms for pions and kaons dominates (first two terms in (\ref{nudyn})).

Figure \ref{nudynppi} shows the measured $p/\pi$ fluctuations in terms of $\nu_{dyn}$ as a function of centrality for Cu+Cu collisions at $\sqrt{s_{NN}}$ = 22.4 GeV (red circles), UrQMD (blue triangles), and HIJING (black squares); scaled by the uncorrected charged particle multiplicity, $dN/d\eta$. Unlike $\nu_{dyn,K/\pi}$, $\nu_{dyn,p/\pi}$ is always negative as a function of centrality in the data. HIJING predicts positive fluctuations, while UrQMD agrees within a factor of two in central collisions, but crosses zero at 30-40\% most central collisions. For the data, correlated production of protons and pions (third term in (\ref{nudyn})) from resonances and particle decays dominates. This causes $\nu_{dyn,p/\pi}$ to be negative. The models have different combinations of resonances, which appear almost totally uncorrelated in HIJING and partially correlated in UrQMD. A linear fit to the data is also shown in Figure \ref{nudynppi}. $\nu_{dyn,p/\pi}$ does not appear to scale with $dN/d\eta$, as $\nu_{dyn,K/\pi}$ does. Though the magnitudes are different, this quantitative effect is reproduced by both HIJING and UrQMD.

\begin{figure}
\centering
\subfigure[]{
\includegraphics[scale=0.37]{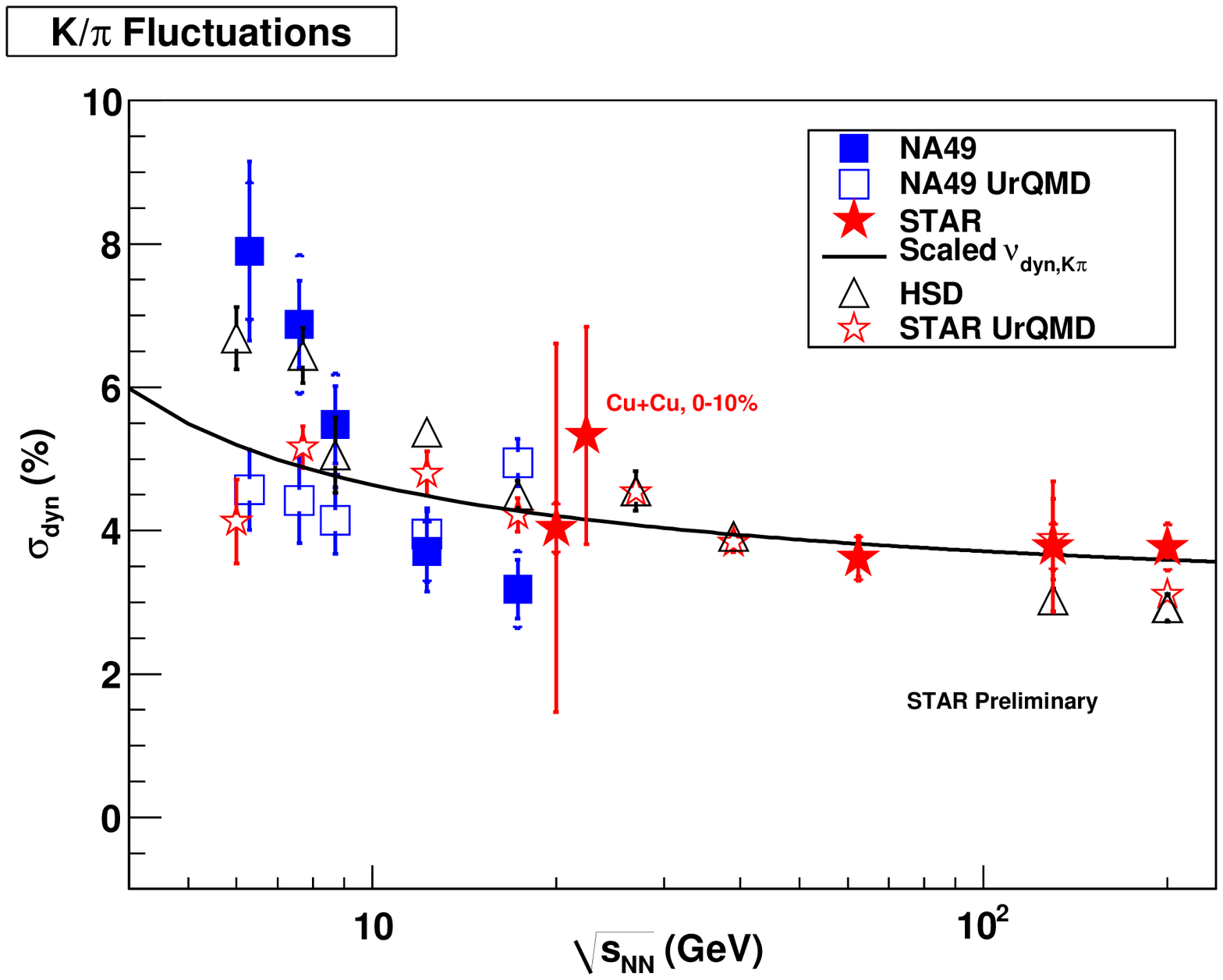} 
\label{kpi_excitation}
}
\subfigure[]{
\includegraphics[scale=0.37]{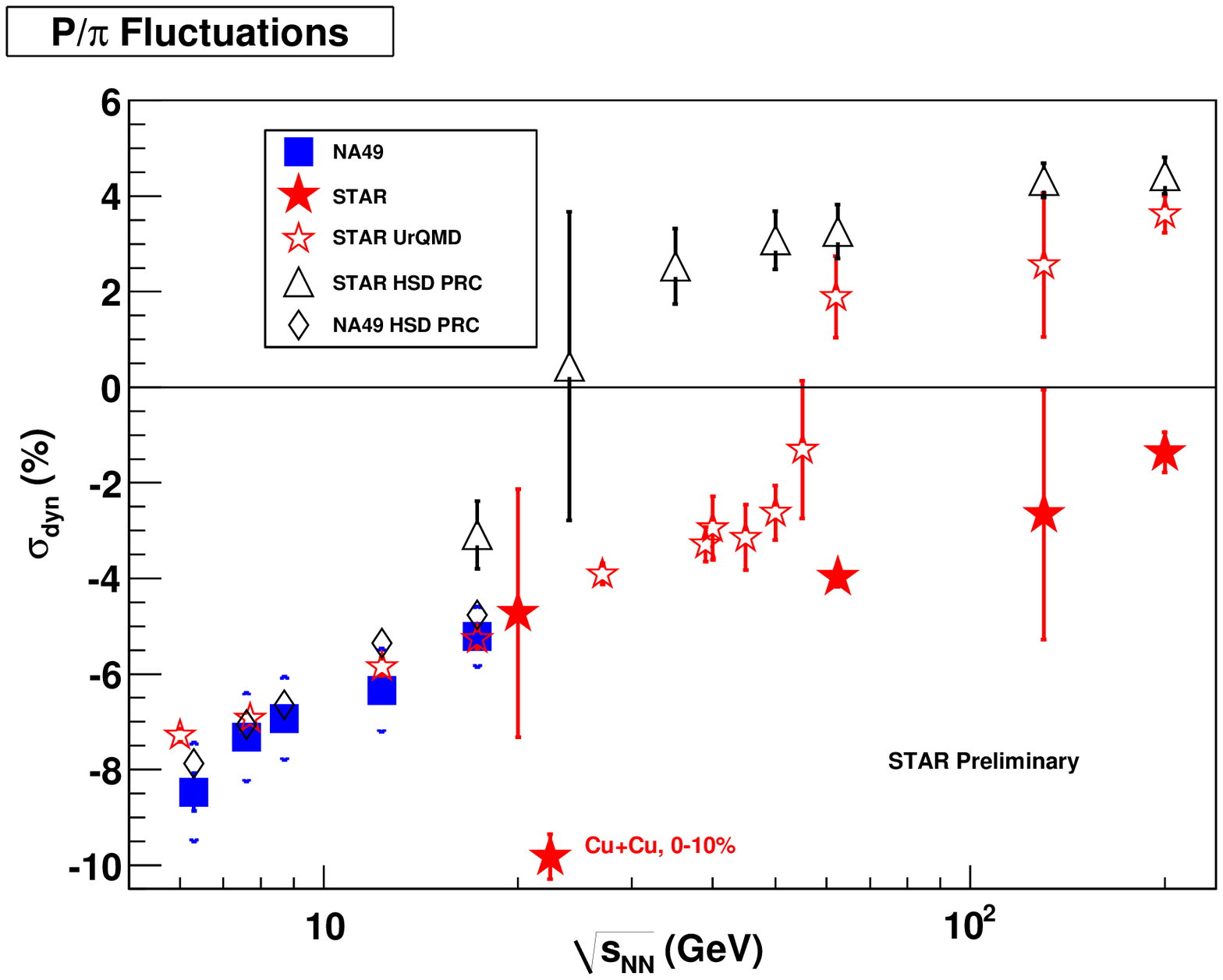} 
\label{ppi_excitation}
}
\caption{$K/\pi$ (left) and $p/\pi$ (right) fluctuations expressed as $\sigma_{dyn}$, as a function of incident energy. Data is shown from both the NA49 (solid blue squares) and STAR experiments (solid red stars) from central collisions, Pb+Pb, 0-3.5\% for NA49 and Au+Au, 0-5\% for STAR, except for the Cu+Cu point at $\sqrt{s_{NN}}$ = 22.4 GeV (0-10\%). Also shown are model calculations from UrQMD using the NA49 (open blue squares) and STAR (open red stars) experimental acceptances and HSD (open black triangles) with the STAR and NA49 (open black diamonds) acceptance. Models labeled ``STAR'' are Au+Au collisions, while models labeled ``NA49'' are Pb+Pb. The solid black line in Figure \ref{kpi_excitation} shows the relationship between $\nu_{dyn}$ at high energies converted to $\sigma_{dyn}$ and extrapolated to lower energies. Errors include both statistical and systematic effects, except for the STAR point in Cu+Cu at $\sqrt{s_{NN}}$ = 22.4 GeV, which has only statistical errors.}
\label{}
\end{figure}

The excitation function for $K/\pi$ fluctuations expressed as $\sigma_{dyn}$ is shown in Figure \ref{kpi_excitation}. To convert $\nu_{dyn}$ to $\sigma_{dyn}$, the relationship in (\ref{signudyn}) was used. Errors were also propagated from $\nu_{dyn}$ to $\sigma_{dyn}$. There is a strong decrease with increasing incident energy for the NA49, Pb+Pb 0-3.5\% results (solid blue squares). The STAR results (solid red stars) for 0-5\% Au+Au at $\sqrt{s_{NN}}$ = 19.6, 62.4, 130, and 200 GeV, and 0-10\% Cu+Cu at $\sqrt{s_{NN}}$ = 22.4 GeV all show the same value of $\sigma_{dyn}$ within errors. $\sigma_{dyn}$ is approximately flat above $\sqrt{s_{NN}}$ = 10 GeV for all systems. Also shown are model predictions for UrQMD processed for the NA49 (open blue squares) and STAR (open red stars) experimental acceptances, respectively. A third model prediction, Hadron String Dynamics (HSD) \cite {HSD} processed using the STAR experimental acceptance (open black triangles) is also plotted.

\begin{figure}
\centering
\includegraphics[scale=0.37]{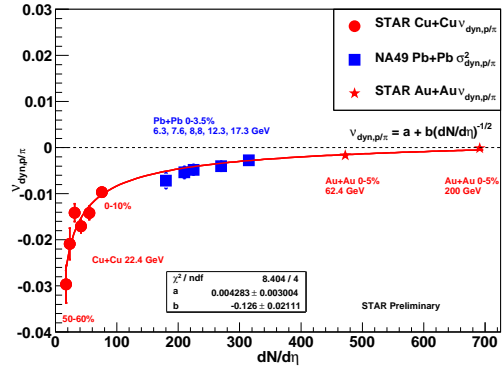}
\caption{$p/\pi$ fluctuations expressed as $\sigma_{dyn}$, as a function of incident energy. Data is shown from both the NA49 (solid blue squares) and STAR experiments (solid red stars) from central collisions, Pb+Pb, 0-3.5\% for NA49 and Au+Au, 0-5\% for STAR, except for the Cu+Cu point at $\sqrt{s_{NN}}$ = 22.4 GeV (0-10\%). Also shown are model calculations from UrQMD using the NA49 (open blue squares) and STAR (open blue circles) experimental acceptances and HSD (open black triangles) with the STAR and NA49 (open black diamonds) acceptances. Errors include both statistical and systematic effects, except for the STAR data points, which contain only statistical errors.}
\label{ppi_syssize}
\end{figure}

Figure \ref{ppi_excitation} shows the excitation function for $p/\pi$ fluctuations expressed using the variable $\sigma_{dyn}$. The general trend demonstrated by the data is negative fluctuations that increase positively with increasing incident energy for both NA49, Pb+Pb 0-3.5\% results (solid blue squares) and STAR results (solid red stars) for 0-5\% Au+Au collisions. The outlier data point is from 0-10\% Cu+Cu collisions at $\sqrt{s_{NN}}$ = 22.4 GeV, which is much more strongly negative. Both UrQMD with the STAR experimental acceptance (open red stars), and HSD with the NA49 and STAR experimental acceptances (open diamonds and triangles, respectively, reproduce most of the low energy data and over predict the high energy data points. To quantify the effect of system size on $p/\pi$ fluctuations, Figure \ref{ppi_syssize} shows $\nu_{dyn,p/\pi}$ as a function of charged particle multiplicity (centrality), $dN/d\eta$. Plotted are 0-10\% to 50-60\% Cu+Cu collisions at $\sqrt{s_{NN}}$ = 22.4 GeV (red circles), the most central 0-3.5\% Pb+Pb collisions at $\sqrt{s_{NN}}$ = 6.3, 7.6, 8.8, 12.3, and 17.3 GeV from NA49 (blue squares), and the most central 0-5\% Au+Au collisions at $\sqrt{s_{NN}}$ = 62.4 and 200 GeV from STAR (red stars). The data points from NA49 were converted to $\nu_{dyn}$ by plotting $\sigma_{dyn}^{2}$. $\nu_{dyn,p/\pi}$ displays a strong system size dependence for small $dN/d\eta$. A fit of the form $\nu_{dyn,p/\pi} = a + b(dN/d\eta)^{-1/2}$ was made to only the Cu+Cu data at $\sqrt{s_{NN}}$ = 22.4 GeV and extrapolated to the higher energy data points. The data from both STAR and NA49 is well described by the extrapolated fit. Further study is necessary to fully understand the physical implications of this particular fit form.

\section{Summary}

The first results on particle ratio ($K/\pi$ and $p/\pi$) fluctuations from Cu+Cu collisions at $\sqrt{s_{NN}}$ = 22.4 GeV has been presented. The most central 0-10\% result for $K/\pi$ fluctuations from Cu+Cu collisions at $\sqrt{s_{NN}}$ = 22.4 GeV is in agreement with the next closest STAR data point, 0-5\% most central Au+Au collisions at $\sqrt{s_{NN}}$ = 19.6 GeV. However, $p/\pi$ fluctuations from the most central 0-10\% result from Cu+Cu collisions at $\sqrt{s_{NN}}$ = 22.4 GeV has a larger negative value than 0-5\% most central Au+Au collisions at $\sqrt{s_{NN}}$ = 19.6 GeV and 0-3.5\% most central Pb+Pb collisions at $\sqrt{s_{NN}}$ = 17.3 GeV from the NA49 experiment. This has been quantified by demonstrating the strong system size dependence of $p/\pi$ fluctuations. It has also been shown that $\nu_{dyn,K/\pi}$ scales linearly with $dN/d\eta$ for data and models in Cu+Cu at $\sqrt{s_{NN}}$ = 22.4 GeV, while $\nu_{dyn,p/\pi}$ does not. Particle ratio fluctuations will be important to study at all energies in the proposed RHIC Critical Point Search. This study provides a baseline for future measurements of $K/\pi$ and $p/\pi$ fluctuations.
\\
\\

\bibliography{all}
\bibliographystyle{unsrt}

\end{document}